\documentclass[aps,prl,twocolumn,amsmath]{revtex4}
\usepackage[dvips]{graphicx}
\usepackage{amssymb,amsfonts,amsmath}
\usepackage{color}

\begin{document}



\author{Vasili Perebeinos} \email{vperebe@us.ibm.com}
\affiliation{IBM Research Division, T. J. Watson Research Center,
Yorktown Heights, New York 10598}
\author{Phaedon Avouris} 
\affiliation{IBM Research Division, T. J. Watson Research Center,
Yorktown Heights, New York 10598}

\title{Current Saturation and Surface Polar Phonon Scattering in Graphene}
\date{\today}


\begin{abstract}

We present a study of transport in graphene devices on polar insulating substrates using a tight-binding model. The mobility is computed using a multiband Boltzmann treatment. We provide the scaling of the surface polar phonon contribution to the low-field mobility with carrier density, temperature, and distance from the substrate. At high bias, we find that graphene self-heating effect is essential to account for the observed saturated current behavior. We predict that by optimizing the device cooling, the high bias currents can be significantly enhanced.
\end{abstract}


\maketitle

The excellent transport and optical properties of graphene \cite{GeimRMP} provide strong motivation for research into possible applications of this
material in nanoscale electronics and optoelectronics \cite{GeimScience09,avouris-dev,Xia}. The electrostatic modulation of the graphene channel through gates yields very promising two-dimensional field-effect devices for analog and radio-frequency applications \cite{YuMingRF,ShepardRF}.
Such devices should ideally be operated in the saturation limit \cite{Shepard}. Indeed, recently it has been shown that the current saturates as the source-drain field is
increased to a few Volts per micron \cite{Shepard,Freitag,Barreiro}. While elastic scattering determines the rate at which current increases with the applied bias,
the current saturation process has been attributed to either the inelastic scattering of electrons by surface polar phonons (SPP) in the polar substrates \cite{Shepard,Freitag} or the intrinsic graphene optical phonons \cite{Barreiro}.
Significant heating of the graphene devices operated under high bias conditions
has also been observed by Raman spectroscopy \cite{Freitag}. However, theoretical studies of such temperature effects on the inelastic scattering and device performance optimization by substrate engineering and thermal management are lacking.

In this Letter, we evaluate diffusive transport properties of graphene on SiO$_2$, HfO$_2$, and SiC polar substrates by solving the Boltzmann transport equation (BTE) in the presence of both intrinsic graphene phonons and substrate SPP phonons. The electronic structure of graphene is described by a $\pi$-orbital tight-binding model with a hoping parameter $t_0=3.1$ eV, which gives a Fermi velocity $v_F=(\sqrt{3}/2)t_0a/\hbar\approx 10^6$ m/s, where $a=0.246$ nm is the graphene lattice constant. For the electron-phonon scattering we use the Su-Schrieffer-Heeger (SSH) model \cite{SSH} for modulation of the $\pi$-orbital overlap $t=t_0-g\delta R_{CC}$  with $g=5.3$ eV/\AA \ as used in the mobility calculations in carbon nanotubes \cite{PerebeinosCNT1}. The parameters for SPP scattering in different substrates
are given in Table~\ref{tab1}.

\begin{figure}[h!]
\includegraphics[height=4.00in]{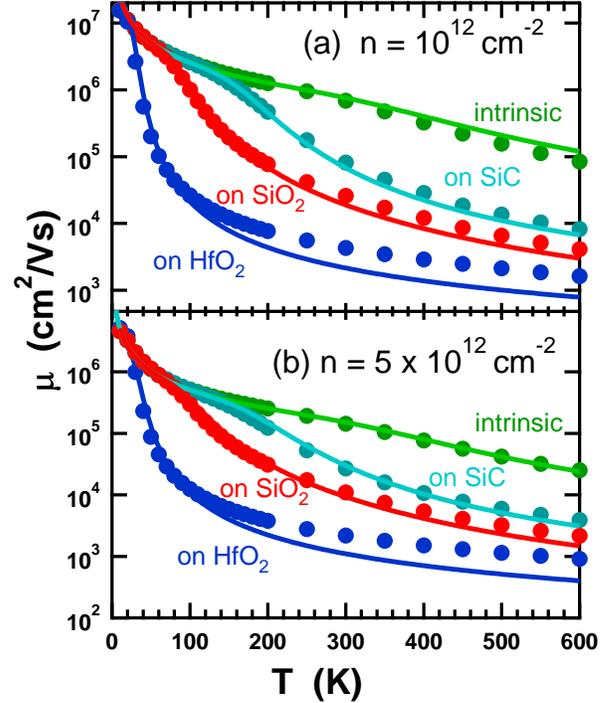}
\caption{\label{Fig1} (Color online) Temperature dependence of the low-field mobility in graphene on different substrates: green - intrinsic, cyan - on SiC, red - on SiO$_2$, blue - on HfO$_2$ at carrier density (a) $n=10^{12}$ cm$^{-2}$ and (b) $n=5 \times 10^{12}$ cm$^{-2}$. The solid curves are fits to Eq. (\protect{\ref{eq10}}) with three global fit parameters: characteristic sound velocity $v_{ph}=17.3$ km/s, energy of the optical phonon in graphene $\hbar\omega_{op}=0.19$ eV, and $\beta\approx 0.0115$ in Eq. (\protect{\ref{eq9}}).}
\end{figure}

The low-field mobility in pristine graphene, in the absence of charged impurities and defects, is determined by scattering from the graphene phonons.  The results for the low-field mobility are shown in Fig.~\ref{Fig1}. Within the SSH model there are two transverse (TA) and longitudinal (LA) acoustic phonon branches whose electron-phonon couplings can be approximated by the deformation potential \cite{Rice,Mahan}: $D_{ac}=3ga\kappa/(4\sqrt{3})$, where the reduction factor $\kappa=v_{TA}^2/(\sqrt{2}(v_{LA}^2-v_{TA}^2))$ was introduced in \cite{Mahan}. For the valence phonon model \cite{PerebeinosTersoff} used here we obtain $D_{ac}\approx 2.6$ eV. The TA and LA modes have different angle dependencies for the electron-phonon couplings $\vert M_{k,k+q}\vert^2=D_{ac}^2q^2\hbar/(2M_C\omega_q)(1\pm\cos(3(\theta_k+\theta_{k+q})))/2$ \cite{Rice}, where $\theta_k$ is given by $\cos \theta_{k}=\delta k_x/\vert \delta k\vert$ and $\sin\theta_{k}=\delta k_y/\vert \delta k\vert$ for the carrier wavevector $k$ in the vicinity of the $K$-point, i.e. $k=K+\delta k$. Thus, the acoustic phonon mobility contribution can be approximated by an angle independent coupling and a characteristic sound velocity $v_{ph}$. The low-field mobility in this model has been calculated \cite{DasSarma} and in the large temperature limit it was found to be \cite{Ferry}:
\begin{eqnarray}
\mu_{ac}=\frac{e\rho_m\hbar v_F^2v_{ph}^2}{2\pi D_{ac}^2}\frac{1}{nk_BT}
\label{eq7}
\end{eqnarray}
where $\rho_m$ is graphene mass density and $n$ is the carrier concentration. 

The two optical phonons at $\Gamma$ point have couplings $\vert M_{k,k+q}^{s,s'}\vert^2=D_{\Gamma}^2\hbar/(2M_C\omega_{\Gamma})(1\pm ss'\cos(\theta_k-\theta_{k+q}))$ for LO (- sign) and TO (+ sign) modes respectively \cite{Piscanec,Ando1}, where $D_{\Gamma}=3g/\sqrt{2}\approx 11.2$ eV/\AA \ \cite{Ando2}, $s=1$ for electrons and $s=-1$ for holes. The K-point TO phonon mode has an electron-phonon coupling twice as large \cite{PerebeinosCNT1,Piscanec} with the angle dependence given by
$\vert M_{k,k+q}^{s,s'}\vert^2=D_{\Gamma}^2\hbar/(M_C\omega_{K})(1-ss'\cos(\theta_k-\theta_{k+q}))$ \cite{Piscanec}. The effect of the optical phonons (both at $\Gamma$ and $K$) on the low-field mobility can be calculated according to \cite{Ferry}:
\begin{eqnarray}
\mu_{op}=\frac{e\rho_m v_F^2 \omega_{op}}{2\pi D_{op}^2}\frac{1}{nN_{op}}
\label{eq8}
\end{eqnarray}
where $N_{op}$ is the Bose-Einstein occupation number of optical phonons and $D_{op}=2D_{\Gamma}=22.4$ eV/\AA  \  is an effective electron-optical phonon coupling \cite{NoteKep} .

\begin{table}[hb]
\caption{\label{tab1} Parameters for the SPP scattering for graphene on SiO$_2$, HfO$_2$, and SiC substrates. The surface phonon (SO)
frequencies are obtained from the bulk longitudinal (LO) phonons
as  $\omega_{SO}=\omega_{LO}\left(\frac{1+1/\epsilon_{0}}{1+1/\epsilon_{\infty}}\right)^{1/2}$}
\begin{ruledtabular}
\begin{tabular}{cccc}
 & SiO$_2$\cite{SiO2param} & HfO$_2$\cite{Fischetti} & SiC\cite{SiCbook} \\
\hline
$\varepsilon_0$ &  3.9 & 22.0  & 9.7 \\
$\varepsilon_{i}$ & 3.36 & 6.58 & - \\
$\varepsilon_{\infty}$ & 2.40 & 5.03 & 6.5 \\
$\hbar\omega_{SO1}$ in meV & 58.9 & 21.6 & - \\
$\hbar\omega_{SO2}$ in meV & 156.4 & 54.2 & 116.0 \\
$F_1^2$ in meV & 0.237 & 0.304 & - \\
$F_2^2$ in meV & 1.612 & 0.293 & 0.735 \\
\end{tabular}
\end{ruledtabular}
\end{table}

The SPP scattering affects the temperature dependence of the mobility in graphene \cite{FuhrerSPP,Gunea,Jena} and carbon nanotubes \cite{PerebeinosSPP} on polar substrates. In graphene it is given by \cite{Gunea,Jena}:
\begin{eqnarray}
&&\vert <\Psi^s_k\vert V_{spp}\vert\Psi^{s'}_{k+q}>\vert^2=
\nonumber \\
&=&\frac{1+ss'\cos(\theta_{k+q}-\theta_k)}{2}\frac{4\pi^2 e^2F_{\nu}^2}{Aq}e^{-2qz_0}
\label{eq6}
\end{eqnarray}
where $z_0\approx 3.5$  \AA \ is the van der Waals distance between the polar substrate and the graphene flake \cite{NoteAngle}. The magnitude of the polarization field is
given by the Fr${\rm \ddot{o}}$hlich coupling:
$F^2_{\nu}=\frac{\hbar\omega_{SO,\nu}}{2\pi}\left(\frac{1}{\varepsilon_{\infty}+1}-\frac{1}{\varepsilon_{0}+1}\right)$,
where $\hbar\omega_{SO,\nu}$ is a surface phonon energy and
$\varepsilon_0$ and $\varepsilon_{\infty}$ are the low- and high-frequency
dielectric constants of the polar substrate. The dielectric constant of air is one. When there are several
SPP phonon modes with an appreciable coupling, then the low- and high-
frequency $\varepsilon$ are understood as an intermediate dielectric functions at
$\omega_i\ll\omega_{SO,\nu}$ for $\epsilon_{0}$ and at $\omega_i\gg\omega_{SO,\nu}$ for
$\varepsilon_{\infty}$ \cite{Fischetti}.

We find that the SPP contribution to the low-field mobility can be approximated as:
\begin{eqnarray}
\mu_{spp, \nu}\approx \beta \frac{\hbar v_F}{e^2}\frac{ev_F}{F_{\nu}^2}\frac{\exp{\left(k_0z_0\right)}}{N_{spp, \nu}\sqrt{ n}}
\label{eq9}
\end{eqnarray}
which is  a non-monotonic function of carrier density $n$. Here $k_0\approx\sqrt{(2\omega_{SPP}/v_F)^2+\alpha n}$, where value of
$\alpha\approx10.5$ was determined from the calculated low-field mobility dependence on $z_0$ (not shown) and parameter $\beta\approx 0.0115$ is a
global fit parameter used in Fig.~\ref{Fig1}. $N_{spp, \nu}$ is the occupation number of the SPP phonons.

The calculated low-field mobility from the BTE solution in Fig.~\ref{Fig1} can be well  fitted using Matthiessen's rule:
\begin{eqnarray}
\mu^{-1}=\mu_{ac}^{-1}+\mu_{op}^{-1}+\sum_{nu}\mu_{spp,\nu}^{-1}
\label{eq10}
\end{eqnarray}
where mobility contributions due to the acoustic, optical, and SPP phonons are given by Eq.~(\ref{eq7}), (\ref{eq8}), and (\ref{eq9}) respectively. Eq.~(\ref{eq10}) describes BTE results remarkably well except for the case of HfO$_2$ at high temperatures, where both SPP phonons have similar coupling strengths.

\begin{figure}[h!]
\includegraphics[height=5.0in]{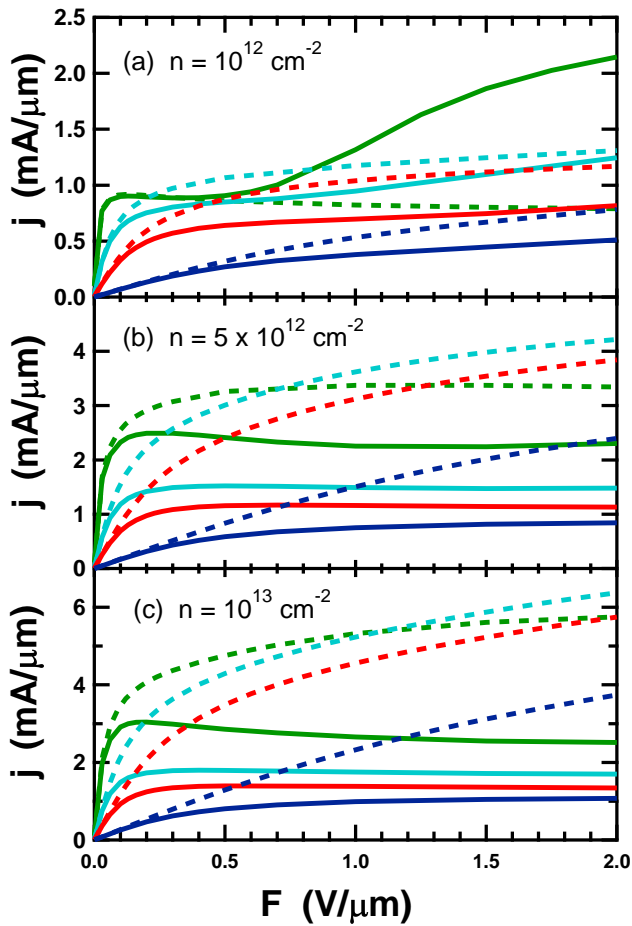}
\caption{\label{Fig2} (Color online) Current density electric field dependence in graphene on different polar substrates: green - intrinsic, cyan - on SiC, red - on SiO$_2$, blue - on HfO$_2$ at carrier density (a) $n=10^{12}$ cm$^{-2}$, (b) $n=5 \times 10^{12}$ cm$^{-2}$, and (c) $n=10^{13}$ cm$^{-2}$. The dashed curves show calculations for phonons at $T_{amb}=300 $ K and solid curves show self consistent calculations with $T=T_{amb}+j(T)F/g$ for $g=0.278$ kW/(K cm$^2$) \cite{Freitag}.}
\end{figure}

At high bias, the transport is typically described by the saturated current model \cite{Shepard,PerebeinosCNT1}:
$j=\frac{\sigma F}{1+\sigma F/j_{sat}}$, where $\sigma=en\mu$ is a low-field conductivity and $j_{sat}$ is a saturation current. In the full saturation regime only carriers around Fermi energy $E_{F}$ in the energy window $E_F\pm\hbar\Omega/2$ contribute to the current. The saturated current can be readily calculated \cite{Shepard,Freitag,Barreiro} and for $E_F>\hbar\Omega/2$:
\begin{eqnarray}
j_{sat}\approx\frac{2e}{\pi^2}\Omega\frac{E_F}{\hbar v_F}
\label{eq12}
\end{eqnarray}
Here $\Omega$ is a characteristic frequency of the phonon responsible for the current saturation.

The current densities as a function of electric field are shown in Fig.~\ref{Fig2} for graphene on different substrates. When phonons are kept in thermal equilibrium at $T_{amb}=300$ K, the current does not show full saturation for the experimentally relevant source-drain fields up to 2 V/$\mu$m.
At low densities, the current shows negative differential conductance for scattering by intrinsic  graphene phonons because of the deviation of the bandstructure from the linear band dispersion, similar to the effect of the non-parabolicity in carbon nanotubes \cite{PerebeinosCNT1}. The current at high bias (2 V/$\mu$m) in Fig.~\ref{Fig1} is proportional to $E_F$ as suggested by equation Eq.~(\ref{eq12}). However, the values of $\hbar\Omega\approx 259, 238, 150$ meV for SiC, SiO$_2$, and HfO$_2$ substrates correspondingly are significantly larger than the SPP phonon energies in Table~\ref{tab1}.

Recently, it has been shown \cite{Freitag} that graphene under high bias can experience a significant self-heating with graphene phonon temperatures reaching up to 1000 K. The temperature was found to be proportional to the Joule losses $T=T_{amb}+jF/g$, where the ambient temperature was about $T_{amb}=300$ K and $g=0.278$ kW/(K cm$^2$) \cite{Freitag}. In the presence of the SPP scattering, electrons can give their energy directly to the substrate SPP phonons \cite{Rotkin}, which can be heated, in principle, to temperatures higher than that of the graphene phonons. In our ``self-heating model'' we   assume that SPP and graphene phonons are heated to the same temperature which is proportional to the Joule losses found self-consistently, i.e. $j=j(T)$. As a result of self-heating the current densities drop by up to a factor of four at high biases, especially at high carrier densities, as shown in Fig.~\ref{Fig2}. Moreover, the current shows true saturation at experimentally accessible source-drain fields. The high bias currents are still proportional to the $E_F$ in the self-heating model; however, the current does not extrapolate to zero at low density. The values of $\hbar\Omega\approx 61, 49, 39$ meV for SiC, SiO$_2$, and HfO$_2$ respectively, are extracted from the current values at 2V/$\mu$m using Eq.~(\ref{eq12}) and are comparable to $\hbar\Omega$ reported in \cite{Shepard,Freitag}. In the self-heating model, the temperature increase contributes to the pinch-off effect at low carrier density (see Fig.~\ref{Fig1}a) as was recently observed in Ref.~\cite{Shepard}. At high density, the self-heating with intrinsic graphene phonons is predicted here to lead to negative differential conductance similar to the effect observed in carbon nanotubes \cite{Dai}.

\begin{figure}[h!]
\includegraphics[height=3.60in]{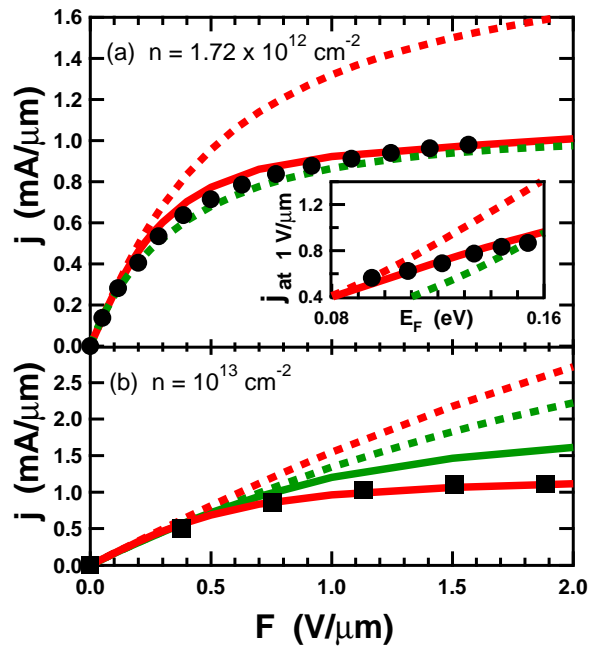}
\caption{\label{Fig3} (Color online) Modeling of the current-voltage characteristics measured in Ref.~\cite{Freitag} and \cite{Barreiro} in graphene devices on SiO$_2$ substrates. The green and red dashed curves, correspondingly, are BTE solutions for graphene phonons and both graphene and SPP phonons scattering at $T_{amb}=300$ K including
Coulomb scattering (see text). The solid green and red curves are self-heating model calculations with graphene phonon only and both SPP and graphene phonon scattering, respectively. (a) Black circles are measurements from Ref.~\cite{Barreiro}
(hole branch). The red solid curves used $g=0.455$ kW/(K cm$^2$) as a fit parameter. The inset shows calculated current density at 1V/$\mu$m field compared to the $I_{SR}$ (black solid circles) from Ref.~\cite{Barreiro}. (b) Black squares are from Ref.~\cite{Freitag}. The solid curves are calculated with $g=0.278$ kW/(K cm$^2$) \cite{Freitag}, which are essentially zero parameter fits.}
\end{figure}

In the experiment, the low-field mobilities can be significantly effected by the presence of defect and charge impurity scattering \cite{Fuhrer,DasSarma2}. As in Ref.~\cite{Barreiro} we include charge impurity scattering in the modeling of the experimental I-V characteristics in Fig.~\ref{Fig3}, following \cite{DasSarma2}. The observed mobility of
about 1000 cm$^2$/Vs in Ref.~\cite{Freitag} can be reproduced in our calculations by assuming scattering with charge impurities of density $n_i=4.5 \times 10^{12}$ cm$^{-2}$ and a smaller $n_i=3.5 \times 10^{12}$ cm$^{-2}$ in the presence of SPP scattering in SiO$_2$ substrate. The carrier density was fixed by the gate at $n\approx 10^{13}$ cm$^{-2}$ in \cite{Freitag} and in Fig.~\ref{Fig3}b we show that the calculated current is significantly larger at high biases in the presence of both intrinsic graphene and SPP phonon scattering if the temperature is fixed at room temperature. Most importantly, the current does not show the saturation that is observed in the experiment. On the other hand, using the experimentally measured temperatures, our self-heating model with $g=0.278$ kW/(K cm$^2$) and including the SPP scattering very nicely reproduces the experiment. At the same time, the self-heating model with only the graphene  phonons active does not show full saturation even at fields up to 2 V/$\mu$m and it overestimates the measured current at high biases.

The four probe I-V characteristics at low carrier density \cite{Barreiro} were analyzed using defect and charge impurity scattering and intrinsic graphene scattering. While the overall agreement between the theory and the experiment was considered satisfactory in  \cite{Barreiro}, here, we show that the agreement can be improved if the data are analyzed within the self-heating model in the presence of SPP scattering from the SiO$_2$ substrate. To reproduce the low-field mobility of about 11000 cm$^2$/Vs in \cite{Barreiro}, we use $n_i=4.3 \times 10^{11}$ cm$^{-2}$ for the Coulomb scattering. A similar value was used in \cite{Barreiro}. The model that includes only intrinsic graphene scattering gives a good agreement with the measurements \cite{Barreiro} at $n\approx 1.72 \times 10^{12}$ cm$^{-2}$ in Fig.~\ref{Fig3}a. However, at lower carrier density the agreement is worse as seen from the inset in Fig.~\ref{Fig3}a and as it was also found by Barreiro et. al. \cite{Barreiro} in Fig.~2d. In the presence of SPP scattering, we use an impurity concentration of  $n_i=2 \times 10^{11}$ cm$^{-2}$ to get the same low-field mobility. While the isothermal calculations overestimate the measured current at high bias, the self-heating model with $g=0.455$ kW/(K cm$^2$) reproduces the I-V characteristics at a fixed density in  Fig.~\ref{Fig3}a fairly well. Moreover, the agreement of the self-heating model with experiment holds even at low carrier densities as can be seen in the inset of Fig.~\ref{Fig3}a, which shows the experimentally measured current at minimum $dI/dV$ \cite{Barreiro} compared to the calculated current densities at 1 V/$\mu$m.

In conclusion, our calculations suggest that SPP scattering is the likely mechanism for the current saturation and that the observed full current saturation can only be accounted by the self-heating model. Without self-heating, the current densities are predicted to be too high for either graphene phonon scattering or SPP scattering. Therefore, saturated currents can be enhanced if efficient device cooling is applied by the appropriate choice of substrate and the optimization of the graphene/substrate contact thermal resistance.

We gratefully acknowledge M.~Freitag for providing data for Fig.~\ref{Fig3}b and I.~Meric for helpful discussions.


\begin{thebibliography}{99}
\bibitem{NoteKep} The angle integration in the scattering rate for $K$-point phonons gives a factor of $3/2$ larger contribution than that for $\Gamma$-point
phonons scattering.
\bibitem{NoteAngle} For wavevector $k$ arbitrary away from $K$-point, the angle $\theta_k$ in Eq.~(\ref{eq6}) is given by  $\exp{(i\theta_k)}=\gamma_k/\vert \gamma_k \vert$, where $\gamma_k=\exp{\left(-ik_ya/\sqrt{3}\right)}+2\cos{\left(k_xa/2\right)}\exp{\left(ik_ya/(2\sqrt{3})\right)}$.
\bibitem{SiO2param} We use some avaraged values from several reports \cite{Fischetti,Gunea,Jena}.
\end{thebibliography}
\end{document}